\documentclass[12pt]{article}
\usepackage{amssymb}
\usepackage{amsmath}
\usepackage{epsfig}
\usepackage{graphicx}
\usepackage{wrapfig}

\begin{document}
\newcommand{\rrtth}{$\gamma\gamma \to t \bar t h^0$ }
\newcommand{\eetth}{$e^+ e^- \to t \bar t h^0$ }

\title{ The effects of large extra dimensions on associated $t\bar{t} h^0$ production
at linear colliders \footnote{Supported by National Natural
Science Foundation of China.}} \vspace{3mm}

\author{\small{ Sun Hao$^{2}$, Zhang Ren-You$^{2}$, Zhou Pei-Jun$^{2}$, Ma Wen-Gan$^{1,2}$,
Jiang Yi$^{2}$, Han Liang$^{2}$ }\\
{\small $^{1}$ CCAST (World Laboratory), P.O.Box 8730, Beijing 100080, P.R.China} \\
{\small $^{2}$ Department of Modern Physics, University of Science and Technology}\\
{\small of China (USTC), Hefei, Anhui 230027, P.R.China}  }

\date{}
\maketitle \vskip 12mm

\begin{abstract}
In the framework of the large extra dimensions (LED) model, the
effects of LED on the processes \rrtth and \eetth at future linear
colliders are investigated in both polarized and unpolarized
collision modes. The results show that the virtual Kaluza-Klein
(KK) graviton exchange can significantly modify the standard model
expectations for these processes with certain polarizations of
initial states. The process \rrtth with $\sqrt{s}=3.5~TeV$ allows
the effective scale $\Lambda_T$ to be probed up to 7.8 and 8.6 TeV
in the unpolarized and $P_{\gamma} = 0.9$, J=2 polarized $\gamma
\gamma$ collision modes, respectively. For the \eetth process with
$\sqrt{s}=3.5~TeV$, the upper limits of $\Lambda_T$ to be observed
can be 6.7 and 7.0 TeV in the unpolarized and $P_{e^+} = 0.6$,
$P_{e^-} = 0.8$, $-$$+$ polarized $e^+e^-$ collision modes,
respectively. We find the \rrtth channel in $J=2$ polarized photon
collision mode provides a possibility to improve the sensitivity
to the graviton tower exchange.
\end{abstract}

\vskip 3cm


\vfill \eject

\baselineskip=0.36in

\renewcommand{\theequation}{\arabic{section}.\arabic{equation}}
\renewcommand{\thesection}{\Roman{section}}
\newcommand{\nb}{\nonumber}

\makeatletter      
\@addtoreset{equation}{section}
\makeatother       

\section{Introduction}
\par
In the late 1990's, a new solution to the hierarchy problem was
proposed, which was accomplished by the presence of new large
extra dimensions $\cite{ADD model}$ $\cite{RS model}$, instead of
low-energy supersymmetry $\cite{mssm-nilles}$ or technicolor
$\cite{technicolor}$. Furthermore, the universe might have more
than three spatial dimensions is not a brand new idea, and string
theory has suggested that there could be up to seven additional
spatial dimensions. In the large extra dimensions (LED) model, the
relationship between extra dimensions' number $n$ and LED
compactification radius $R$ is expressed by $\cite{ADD model}$
\begin{eqnarray}
R \sim 10^{30/n - 17} {\rm cm} \times \left(\frac{\rm 1 TeV}{
m_{\rm EW}}\right).
\end{eqnarray}
The case of $n = 1$ is obviously ruled out, since it would modify
Newton's law of gravity at solar-system distances. The case of $n
= 2$ is also likely to be ruled out because of the results from
the gravity experiments at submillimetre distance $\cite{grav
exp1}$, and cosmological constraints from supernova cooling and
distortion of cosmic diffuse gamma radiation as well $\cite{grav
exp2}$. As $n$ increases from 2 to 10, $1/R$ increases from about
$10^{-3}$ eV to about 1 GeV. Therefore, the standard model (SM)
fields (gauge and matter fields) must be confined to the ordinary
4-dimensional space-time manifold, since the standard model has
been tested up to $\sim 10^2$ GeV.

\par
The large extra dimensions model becomes an attractive extension
of the SM because of its possible testable consequences. As
Arkani-Hamed, Dimopoulos, and Dvali $\cite{ADD model}$ proposed,
the SM particles exist in the usual $(3+1)$-dimensional space, and
graviton can propagate in a higher-dimensional space. Another
manifestation of the large extra dimensions model is the existence
of a Kaluza-Klein (KK) tower of massive gravitons which can
interact with the SM fields on the wall $\cite{indirect1}$.
Experimental reviews on probing large extra dimensions are
presented in Refs. $\cite{ED d0}$ $\cite{ED cdf}$ $\cite{ED lhc}$,
$\cite{ED hera}$, $\cite{re-1}$ and $\cite{re-2}$.

\par
Many papers have been concentrated on studying the LED effects on
the processes at high energy colliders. The effective interactions
of graviton and ordinary matter fields (fermions, gauge bosons and
scalars) are presented in Ref. $\cite{Han T}$. There are two
classes of processes to probe LED effects: the real graviton
emission processes and the virtual graviton exchange processes.
For the second class of processes, any significant deviation from
the SM prediction can be considered as the possible signal of the
LED physics. In reference $\cite{lamT}$, the LED signatures of
$\gamma + \not \! \! E$ and $jet + \not  \! \! E$ as well as the
virtual graviton exchange process $f \bar{f} \to \gamma \gamma$ at
$e^+e^-$ and $\mu^+\mu^-$ colliders, are investigated. Reference
$\cite{rrww}$ studied the contributions of the graviton exchange
in the process $\gamma \gamma \to W^+ W^-$. It was found that the
differential cross section as well as the polarization of $W$'s in
the final state are quite sensitive to graviton exchange
especially for certain initial photon polarizations. The exchange
of Kaluza-Klein towers of massive gravitons in fermion pair
production in $e^+e^-$ annihilation and in Drell-Yan production
are studied in Ref. $\cite{indirect1}$. It is concluded that
future linear colliders and the LHC can exclude a string scale up
to several TeV by measuring the LED effects in the $2 \to 2$
processes $e^+e^- \to f \bar f$, $q \bar q \to l^+l^-$, and $gg
\to l^+l^-$. Recently, the studies of LED effects have been
extended to three-body final states processes, such as $t\bar{t}
h^0$, $h^0h^0Z^0$ and $h^0h^0\gamma$ productions at future
$e^+e^-$ and $\mu^+\mu^-$ colliders $\cite{ED 3f1}$ $\cite{ED
3f2}$ $\cite{ED 3f3}$. The results show that for $n=3$, the
processes of $e^+e^- \to h^0h^0Z^0$, $e^+e^- \to t\bar{t} h^0$ and
$e^+e^- \to h^0h^0\gamma$ at a $\sqrt{s}=3~TeV$ linear collider,
can be used to put limits on the effective string scale $M_s$ up
to $6.6~TeV$, $7.9~TeV$ and $7.4~TeV$, respectively. In the
intermediate Higgs boson mass region, the production mechanism
with Higgs boson radiated from top-quark pair, is specially
important. Due to the coupling strength of the top-quark-Higgs
Yukawa coupling is proportional to the top-quark mass, the top
quark Yukawa coupling $g_{t \bar{t} h}$ is very large and the
cross section of process $e^+e^- \to t \bar{t} h^0$ will be
strongly enhanced. Therefore, the $e^+e^- \to t \bar{t} h^0$
process can be used to probe this coupling.

\par
The GLC, NLC and TESLA $e^+e^-$ linear colliders are designed with
colliding energy from $300~GeV$ up to $1~TeV$, while the CLIC at
CERN is expected operating between $3 \sim 5~TeV$ colliding energy
range. An $e^+e^-$ LC can also be converted to a $\gamma\gamma$
collider. This is achieved by using Compton backscattered photons
in the scattering of intense laser photons on an electron beam
\cite{Com}. The resulting $\gamma\gamma$ center of mass system
(CMS) energy is peaked at about $0.8\sqrt{s}$ for the appropriate
choices of machine parameters. Generally $e^+e^-$ collider has the
advantage that the luminosity is higher than $\gamma\gamma$
collider(for example, $500~fb^{-1}/year$ of $e^+e^-$ collider
against $100~fb^{-1}/year$ of $\gamma\gamma$ one), but the
polarization technique for photon is much simpler than positron.
In Ref. \cite{ChenH}, it is concluded that with careful handling
of appropriate efficiency of b-tagging $\cite{btag1}$
$\cite{btag2}$ $\cite{btag3}$ and constraints of the $W^\pm$, t
and $h^0$ masses, the backgrounds of signal $\gamma\gamma \to
t\bar th^0 \to b\bar b b \bar b W^+W^-$ process would be greatly
reduced. Therefore, the $t \bar{t} h^{0}$ production at linear
colliders in $\gamma \gamma$ collision mode would be another
choice in testing the Yukawa coupling between Higgs boson and top
quarks. At linear colliders (LC) \eetth channel at high colliding
energy can be used to test the existence of the virtual KK
exchange\cite{ED 3f2}. But in both the \eetth and \rrtth
processes, the role of proper initial particle polarization to
improve the sensitivity to graviton tower exchange, hasn't been
investigated until now. Therefore, the LED effects related to the
polarizations of initial states in these processes would be
worthwhile to study.

\par
In this paper we study the indirect LED effects induced by the
virtual KK graviton exchange in the processes \eetth and \rrtth,
and emphasize the role of polarization of the initial states in
improving the sensitivity to graviton exchange. In section 2 and
3, we present the calculations, numerical results and discussion
for both processes. Finally, we give a short summary.

\vskip 5mm \vskip 5mm
\section{Analytical Calculations}
\par
In this section, we present the analytical calculations of the two
processes \rrtth and \eetth involving KK graviton exchanges with
different polarizations of incoming particles.

\subsection{$\gamma \gamma \to t\bar t h^0$ process}

\par
At the tree-level, there are nine Feynman diagrams for the process
\rrtth. In Fig.1 we display only three of them which include KK
graviton exchanges. The SM Feynman diagrams and their specific
calculations in unpolarized photon collision mode can be found in
Refs. \cite{ChenH} \cite{chen}. We denote this process as
\begin{equation}
\gamma(p_1, \lambda_1)+\gamma(p_2, \lambda_2) \to t(k_1, e_1) +
\bar{t}(k_2, e_2) + h^0(k_3),
\end{equation}
where $\lambda_{i}$ and $e_{i}$ are photon and
top-quark/anti-top-quark polarizations, $p_i~(i = 1, 2)$ and
$k_i~(i = 1, 2, 3)$ are the four-momenta of incoming photons and
outgoing top-quark, anti-top-quark and Higgs boson, respectively.
All these four-momenta satisfy the on-shell conditions: $p_1^2 =
p_2^2 = 0$, $k_1^2 = k_2^2 = m_t^2$ and $k_3^2 = m_h^2$. The
center-of-mass energy squared is denoted by $s = (p_1+p_2)^2 = 2
p_1 \cdot p_2$. The total helicity of the $\gamma \gamma$ system
is $J=|\lambda_1 - \lambda_2|$.

\par
In this paper, the amplitude of \rrtth process in the LED model is
calculated under the de Donder gauge for simplicity.
$F(k)^{(m)(m^{\prime})}_{AB,CD}$, the graviton propagator under
the de Donder gauge, can be expressed as $\cite{lamT}$
\begin{eqnarray}
F(k)^{(m)(m^{\prime})}_{AB,CD} = \frac{i}{2}
\frac{\delta_{m,m^{\prime}}}{k^2-M_m^2} \left( g_{AC} g_{BD} +
g_{AD} g_{BC} - \frac{2}{D-2} g_{AB} g_{CD} \right),
\end{eqnarray}
where $D = 4+n$, $M_{m}$ is the mass of m-th KK graviton and
$\left(g_{AB}\right)={\rm diag}\{1, -1,...,-1\}$ is the metric
tensor of the $D$-dimensional flat spacetime manifold. The
amplitude for polarized initial photon beams can be divided into
two parts:
\begin{eqnarray}
\label{amplitude} {\cal M}(\lambda_1,\lambda_2) =
\sum_{e_1,e_2}\left[ {\cal M}^{SM}(\lambda_1,\lambda_2,e_1,e_2) +
{\cal M}^{KK}(\lambda_1,\lambda_2,e_1,e_2)\right],
\end{eqnarray}
where ${\cal M}^{SM}$ is the amplitude contributed by the SM-like
diagrams, and
\begin{eqnarray}
{\cal M}^{KK}(\lambda_1,\lambda_2,e_1,e_2) &=&{\cal
M}^{KK}_1(\lambda_1,\lambda_2,e_1,e_2) +
               {\cal M}^{KK}_2(\lambda_1,\lambda_2,e_1,e_2) \nb\\
                      && +{\cal M}^{KK}_3(\lambda_1,\lambda_2,e_1,e_2)
\end{eqnarray}
is the amplitude related to all the three KK graviton exchanged
diagrams presented in Fig.1. The summation is taken over the spins
of final particles, since we do not consider the polarizations of
top quarks. The analytic expression of ${\cal
M}_i^{KK}(\lambda_1,\lambda_2,e_1,e_2)~ (i = 1, 2, 3)$ , which
correspond to the Feynman diagrams in Figs.1(a), (b) and (c)
respectively, are expressed explicitly as
\begin{eqnarray}
{\cal M}^{KK}_1(\lambda_1,\lambda_2,e_1,e_2) &=& -\frac{1}{2}
                   \bar{u}(k_1,e_1) g_{t t h} v(k_2,e_1)
                   \frac{D(s)}{(k_1+k_2)^2-m_h^2}\,\,
                   \rlap/{\epsilon_{\mu}}(p_1,\lambda_1) \, \rlap/{\epsilon_{\nu}}(p_2,\lambda_2)
                   C_{h h G} C_{\gamma \gamma G}  \nb\\
{\cal M}^{KK}_2(\lambda_1,\lambda_2,e_1,e_2) &=& -\frac{1}{2}
                          \bar{u}(k_1,e_1) g_{t t h}
                          \frac{\rlap/{k}_1+\rlap/{k}_3-m_t}{(k_1+k_3)^2-m_t^2}
                          \, \, C_{t t G} v(k_2,e_2) D(s)\,
                          \rlap/{\epsilon_{\mu}}(p_1,\lambda_1) \, \rlap/{\epsilon_{\nu}}(p_2,\lambda_2)
                          C_{\gamma \gamma G} \nb\\
{\cal M}^{KK}_3(\lambda_1,\lambda_2,e_1,e_2) &=& -\frac{1}{2}
                          \bar{u}(k_1,e_1) C_{t t G}
                          \frac{\rlap/{k}_2+\rlap/{k}_3-m_t}{(k_2+k_3)^2-m_t^2}\,\,
                          g_{t t h} v(k_2,e_2) D(s) \nb \\
                          && \rlap/{\epsilon_{\mu}}(p_1,\lambda_1)\, \rlap/{\epsilon_{\nu}}(p_2,\lambda_2)
                          C_{\gamma \gamma G}
                          \, \, ,
\end{eqnarray}
where $g_{t t h}=-igm_t/(2m_W)$, $C_{\gamma \gamma G}, C_{h h G}$
and $C_{t t G}$ are the relevant coupling constants given
explicitly in Appendix, and $D(s)$ represents the summation of the
KK excitation propagators
\begin{eqnarray}
\label{sum of propagator} D(s) = \frac{1}{\bar{M}_{pl}^2}
             \sum_{m \in \mathbb{Z}^n}
             \frac{1}{s - M_m^2}.
\end{eqnarray}
In the LED theory, $M_{pl}$, the 4-dimensional Plank scale, is no
longer a fundamental quantity, and is derived from the size $R$ of
extra dimension. The relationship between the ordinary reduced
Planck scale $\bar{M}_{pl}(=\frac{M_{pl}}{\sqrt{8\pi}})$, the size
of extra dimensions compactification radius $R$, extra dimensions
number $n$ and the fundamental Plank scale $M_s$ is expressed
as\cite{lamT}
\begin{eqnarray}
\bar{M}_{pl}^2 = R^{n}M_{s}^{n+2}.
\end{eqnarray}
As we know, the real part of the sum in Eq.(\ref{sum of
propagator}) is divergent when $n$, the number of
extra-dimensions, larger than 1. Since the LED model is an
effective model, it is only valid below an effective energy scale.
In phenomenology, the most plausible assumption for $D(s)$ is
\cite{lamT}
\begin{eqnarray}
\label{Ds} D(s) = \lambda \frac{4\pi}{\Lambda_{T}^{4}},
\end{eqnarray}
where $\lambda$ is the sign of D(s). This effective theory can be
used only when $\sqrt{s}$ smaller than the effective energy scale
$\Lambda_T$. In the following numerical calculation, we fix
$\lambda$ to be 1.

\par
Finally, the cross section for the polarized initial photon beams
in the framework of the LED model reads
\begin{eqnarray}
\sigma_{LED}^{pol.}(\lambda_1,\lambda_2) =
\frac{N_c}{2|\vec{k}_1|\sqrt{s}} \int {\rm d} \Phi_3  |{\cal
M}(\lambda_1,\lambda_2)|^{2},
\end{eqnarray}
and the phase-space hypercube element for three final particles
process is defined as
\begin{eqnarray}
{\rm d} \Phi_3 =
   \left( \prod_{i=1}^3 \frac{{\rm d}^3 \vec{k}_i}{(2\pi)^3 2k_i^0} \right)\, (2\pi)^4
   \delta\Biggl(p_1+p_2-\sum_{j=1}^3 k_j\Biggr).
\end{eqnarray}
Then the cross section for the unpolarized collision mode can be
expressed as
\begin{eqnarray}
\sigma_{LED}^{unpol.} = \frac{1}{4}
\sum_{\lambda_1,\lambda_2}\sigma_{LED}^{pol.}(\lambda_1,\lambda_2).
\end{eqnarray}

\subsection{\eetth process}
\par
In Refs.\cite{tth}\cite{tthew}, the process \eetth can be used to
probe the Yukawa coupling in the SM. The LED effects on the
process \eetth in unpolarized $e^+e^-$ collision mode have been
investigated by D. Choudhury, et al\cite{ED 3f2}. It was concluded
that the measurement of \eetth cross section with unpolarized
colliding beams, will allow the effective string scale to be
probed up to $M_s=7.9~TeV$ for $n=3$ and $\sqrt{s}=3~TeV$. In this
subsection we shall calculate the effects from additional
contributions of graviton exchange diagrams in the framework of
the LED model on the cross sections of the \eetth channel with
different polarizations of initial states. The Feynman diagrams
for the process \eetth at the lowest level involving KK graviton
exchanges are depicted in Fig.2. We denote this process as
\begin{eqnarray}
e^+(p_1, \lambda_1)+e^-(p_2, \lambda_2) \to t(k_1, e_1) +
\bar{t}(k_2, e_2) + h^0(k_3),
\end{eqnarray}
where $\lambda_{i}$ and $e_{i}(i=1,2)$ are the positron/electron
and top-quark/anti-top-quark helicities, separately. Analogous to
the case in the \rrtth channel, we divide the amplitude into the
SM-like part ${\cal M}^{SM}$ and KK graviton exchange part ${\cal
M}^{KK}$ which can be easily obtained by using the Feynman rules
presented in Ref.\cite{Han T}. For the summation of the KK
excitation propagators $D(s)$, we also use the assumption as shown
in Eq.(\ref{Ds}) and fix $\lambda$ to be 1. The explicit
analytical expressions corresponding to the three Feynman diagrams
involving graviton exchange in Fig.2, ${\cal
M}_i^{KK}(\lambda_1,\lambda_2,e_1,e_2)~ (i = 1, 2, 3)$, are
presented as
\begin{eqnarray}
{\cal M}^{KK}_1(\lambda_1,\lambda_2,e_1,e_2) &=& -\frac{1}{2}
                   \bar{u}(k_1,e_1) g_{t t h} v(k_2,e_1)
                   \frac{D(s)}{(k_1+k_2)^2-m_h^2}\,\,
                   [\bar{v}(p_1)C_{eeG}u(p_2)]
                   C_{h h G} \nb\\
{\cal M}^{KK}_2(\lambda_1,\lambda_2,e_1,e_2) &=& -\frac{1}{2}
                          \bar{u}(k_1,e_1) g_{t t h}
                          \frac{\rlap/{k}_1+\rlap/{k}_3-m_t}{(k_1+k_3)^2-m_t^2}
                          \, \, C_{t t G} v(k_2,e_2) D(s)\,
                          [\bar{v}(p_1)C_{eeG}u(p_2)] \nb\\
{\cal M}^{KK}_3(\lambda_1,\lambda_2,e_1,e_2) &=& -\frac{1}{2}
                          \bar{u}(k_1,e_1) C_{t t G}
                          \frac{\rlap/{k}_2+\rlap/{k}_3-m_t}{(k_2+k_3)^2-m_t^2}\,\,
                          g_{t t h} v(k_2,e_2) D(s)
                           [\bar{v}(p_1)C_{eeG}u(p_2)]\nb \\
                          \, \,
\end{eqnarray}
where the explicit expressions of $C_{t t G}$ and $C_{eeG}$ can be
found in Appendix. Finally, the cross section for the process
\eetth with polarized initial $e^+e^-$ beams in the framework of
the LED model can be calculated by using Eqs.(2.8-10).

\section{Numerical Results}
\par
In this section, we present some numerical results for both the
\rrtth and \eetth processes. In the numerical calculation, we take
the input parameters as follows \cite{hepdata}
\begin{eqnarray}
& &\alpha^{-1}(m_{Z})=127.918, ~~ m_W = 80.425\,{\rm GeV},~~ m_Z =
91.1876\,{\rm GeV}, ~~
m_t= 178.1\,{\rm GeV} \nb \\
& &\mathcal{L}_{e^{+}e^{-}}=500fb^{-1},~~\mathcal{L}_{\gamma
\gamma}= 100fb^{-1}.
\end{eqnarray}
We take the similar acceptance parameters with those in
GLC\cite{GLC} and assume
\begin{eqnarray}
|\cos\theta_h| < 0.966 ~~{\rm and}~~ |\cos\theta_t| < 0.95,
\end{eqnarray}
where $\theta_h$ and $\theta_t$ are the angles between $h^0$, top
quark and the incoming $e^-$ and one of the colliding photon
beams, respectively.

\subsection{\rrtth}
\par
There are five kinds of polarization collision modes for the
\rrtth channel. They are $+$$-$, $-$$+$, $+$$+$, $-$$-$ and
unpolarized collision modes, separately(e.g., the notation of
$+$$-$ represents the helicities of the two initial photons being
$\lambda_1=+1$ and $\lambda_2=-1$). In Fig.3(a), we present the
cross sections of \rrtth as a function of the c.m.s. energy
$\sqrt{s}$ in different kinds of collision modes. Here we take
$m_h=115~GeV$ and $\Lambda_T=3.5~TeV$. Since the cross sections of
the $+$$-$ and $-$$+$ photon polarizations(J=2) are equal, and
same is true for the cross sections of the $+$$+$ and $-$$-$
photon polarizations(J=0), we only depict the results in the
$+$$-$, $+$$+$ and unpolarized photon collision modes, which are
represented by the dashed, dotted and solid curves, respectively.
As the figure shows that in the energy region where
$\sqrt{s}<0.7TeV$, the two curves of the SM and LED model for
unpolarized or $+$$-$ polarized photon collision are overlapped,
but when the colliding energy goes up from $0.7TeV$ to $3.5TeV$,
each curve branches off into two curves. The lower and upper
curves represent the SM and LED results, respectively. In
Fig.3(a), we see that both the SM and LED results for $+$$+$
collision mode are shown by the same dotted curve, that means the
cross sections in both models in $+$$+$ collision mode are almost
the same. In Fig.3(b), we depict the cross sections and the ratio
between the cross sections in the $+$$-$ and $+$$+$ collision
modes, where the ratio $\Delta$ is defined as
\begin{eqnarray}
\Delta = {\sigma_{+-}}/{\sigma_{++}}. \label{Delta}
\end{eqnarray}
It is obviously that the \rrtth channel in the J=2 polarized
photon collision mode provides the posibility to improve the
sensitivity in probing the LED effects. We will focus on this kind
of polarization collision channel in the following calculation.

\par
Fig.4 exhibits the cross sections for the process \rrtth as the
functions of c.m.s. energy $\sqrt{s}$ for different values of
$\Lambda_T$ and $m_h$. The solid curves represent the SM results,
the dashed, dotted, dash-dotted, dash-dotted-dotted and short
dashed curves represent the cross sections for $\Lambda_T$ = 2.5,
3.5, 4.5, 5.5 and 6.5 TeV, respectively. In the figures of Fig.4,
we take $m_h$ to be 115, 150 and 200 GeV for comparison
respectively. The Figs.4(a), (c) and (e) are for unpolarized
collision mode and Figs.4(b), (d) and (f) are corresponding to the
$+$$-$(J=2) polarized collision mode. Again we can see that the
virtual KK graviton exchange can modify the cross section of the
\rrtth process significantly from its SM prediction in both J=2
polarized and unpolarized photon collision modes. As we expected
that when colliding energy $\sqrt{s}$ is large enough, there is an
obvious enhancement for $\sigma_{LED}$, while the $\sigma_{SM}$
goes down slowly with the increment of $\sqrt{s}$.

\par
In Fig.5, we present the dependence of the cross section of the
\rrtth process in both unpolarized and J=2,
$P_{\gamma}(=\frac{N_+-N_-}{N_++N_-})=0.9$ collision modes, on the
parameter $\Lambda_T$ with $m_h = 115$, 150 and 200 GeV,
respectively. For comparison in different colliding energies, we
take the $\gamma \gamma$ colliding energy $\sqrt{s}$ to be 1, 2
and 3~TeV, which are represented by solid, dashed and dotted
curves, respectively. For each fixed values of $\sqrt{s}$ and
$m_h$, the cross section $\sigma_{LED}$ goes down and approaches
to its corresponding SM result($\sigma_{SM}$) as the increment of
$\Lambda_{T}$. This dependence of the cross section for the \rrtth
process on effective scale $\Lambda_{T}$ typically reflects the
relation of ${\cal M}^{KK} \propto \Lambda_{T}^{-4}$, and the
$t\bar th^0$ associated production process in J=2,
$P_{\gamma}=0.9$ polarized collision mode improves obviously the
sensitivity to the LED effective scale $\Lambda_T$.

\par
As demonstrated in above figures, the virtual KK graviton exchange
in both unpolarized and $J=2$, $P_{\gamma}=0.9$ polarized photon
collision modes can obviously modify the cross section of the
\rrtth process from its SM value, if the large extra dimensions
really exist. Since the large extra dimensions can be probed only
when the deviation of the cross section within the framework of
the LED model from its SM value, $\Delta \sigma$, is large enough,
we assume that the LED effect can and can not be observed, only if
\begin{eqnarray}
\label{upper} \Delta\sigma =\sigma_{LED}-\sigma_{SM} \geq
\frac{5\sqrt{\sigma_{LED}\mathcal{L}}}{\mathcal{L}},
\end{eqnarray}
and
\begin{eqnarray}
\label{lower} \Delta\sigma =\sigma_{LED}-\sigma_{SM} \leq
\frac{3\sqrt{\sigma_{LED}\mathcal{L}}}{\mathcal{L}},
\end{eqnarray}
respectively.

\par
In Figs.6(a),(b) and (c), we show the regions in
$\sqrt{s}-\Lambda_{T}$ parameter space for the process \rrtth with
unpolarized incoming photons, where the LED effect can and cannot
be observed according to above criteria, for $m_h$ = 115, 150 and
200 GeV, respectively. Here we take the integrated luminosity for
$\gamma \gamma$ collider as ${\cal L}={\cal L}_{\gamma \gamma} =
100 fb^{-1}$. The underside grey region is excluded by the
unitarity bounds.

\begin{table}[htbp]
\begin{center}
\begin{tabular}{c|c|c}
\hline
$m_h$ = 115[GeV]& \multicolumn{2}{c}{~~~~~~~~~~~~$\Lambda_{T}$~~~~~~  [GeV]}\\[0mm]
\cline{1-3}
$\sqrt{s}$&~~~~$unpol.$~~~&~~~~$J=2~~ pol.$~~~~\\[0mm]
\cline{2-3}\vspace*{-0ex}[TeV]& ~~$3\sigma$~~,~~~$5\sigma$ ~~& $3\sigma$~~,~~~$5\sigma$~~~ \\
\cline{1-3} \hline
0.6& ~~1488,~~1323~~&1889,~~1642~~ \\[0mm]
1.0& ~~2981,~~2629~~&3569,~~3120~~ \\[0mm]
1.5& ~~4354,~~3840~~&5067,~~4439~~ \\[0mm]
2.0& ~~5568,~~4916~~&6382,~~5598~~ \\[0mm]
2.5& ~~6712,~~5931~~&7620,~~6692~~ \\[0mm]
3.0& ~~7805,~~6898~~&8801,~~7742~~ \\[0mm]
3.5& ~~8868,~~7838~~&9785,~~8620~~ \\[0mm]
\hline
\end{tabular}
\\
\begin{tabular}{c|c|c}
\hline
$m_h$ = 150[GeV]& \multicolumn{2}{c}{~~~~~~~~~~~~$\Lambda_{T}$~~~~~~  [GeV]}\\[0mm]
\cline{1-3}
$\sqrt{s}$&~~~~$unpol.$~~~&~~~~$J=2~~ pol.$~~~~\\[0mm]
\cline{2-3}\vspace*{-0ex}[TeV]& ~~$3\sigma$~~,~~~$5\sigma$ ~~& $3\sigma$~~,~~~$5\sigma$~~~ \\
\cline{1-3} \hline
0.6& ~~1254,~~1120~~&1500,~~1306~~ \\[0mm]
1.0& ~~2781,~~2465~~&3256,~~2848~~ \\[0mm]
1.5& ~~4175,~~3690~~&4766,~~4179~~ \\[0mm]
2.0& ~~5389,~~4769~~&6074,~~5336~~ \\[0mm]
2.5& ~~6525,~~5779~~&7288,~~6416~~ \\[0mm]
3.0& ~~7608,~~6749~~&8458,~~7450~~ \\[0mm]
3.5& ~~8664,~~7686~~&9600,~~8458~~ \\[0mm]
\hline
\end{tabular}
\\
\begin{tabular}{c|c|c}
\hline
$m_h$ = 200[GeV]& \multicolumn{2}{c}{~~~~~~~~~~~~$\Lambda_{T}$~~~~~~  [GeV]}\\[0mm]
\cline{1-3}
$\sqrt{s}$&~~~~$unpol.$~~~&~~~~$J=2~~ pol.$~~~~\\[0mm]
\cline{2-3}\vspace*{-0ex}[TeV]& ~~$3\sigma$~~,~~~$5\sigma$ ~~& $3\sigma$~~,~~~$5\sigma$~~~ \\
\cline{1-3} \hline
0.6& ~~898,~~798~~&1015,~~885~~ \\[0mm]
1.0& ~~2548,~~2262~~&2971,~~2596~~ \\[0mm]
1.5& ~~3941,~~3498~~&4495,~~3953~~ \\[0mm]
2.0& ~~5164,~~4587~~&5827,~~5135~~ \\[0mm]
2.5& ~~6314,~~5612~~&7061,~~6235~~ \\[0mm]
3.0& ~~7401,~~6586~~&8236,~~7280~~ \\[0mm]
3.5& ~~8454,~~7527~~&9375,~~8288~~ \\[0mm]
\hline
\end{tabular}
\caption{\em The dependence of $3\sigma$ exclusion limits and
$5\sigma$ observation limits on $\Lambda_{T}$ and $\sqrt{s}$ for
the \rrtth process in both unpolarized and $P_{\gamma}$=0.9, J=2
polarized collision modes. The three tables correspond to $m_h =
115,~ 150$ and 200 GeV, respectively. }
\end{center}
\end{table}

\par
In Table 1, we present the regions of $3\sigma$ exclusion limits
and corresponding $5\sigma$ observation limits for the \rrtth
process in both unpolarized and J=2 polarized collision modes with
$m_h = 115$, $150$, $200~GeV$ and some typical $\sqrt{s}$ values
of LC, respectively. For the J=2 polarized photon collision mode,
the beam polarization $P_{\gamma}$ is set to be $0.9$. From this
table we can find that the process \rrtth in the $P_{\gamma}$=0.9,
J=2(i.e., $+$$-$, $-$$+$ photon polarizations) polarized photon
collision mode can be used to improve the sensitivity of probing
the LED model effects. For $\sqrt{s}$=3.5 TeV and $m_h$=115 GeV,
the observation limits on $\Lambda_{T}$ can be probed up to 8.6
TeV and the exclusion limits on $\Lambda_{T}$ can be probed up to
9.8 TeV.

\par
Fig.7 exhibits the $\sqrt{s}$ dependence of relative discrepancy
between the cross sections in the LED model and SM for process
\rrtth in unpolarized and $J=2$, $P_{\gamma}=0.9$ polarized
collision modes respectively, where the relative discrepancy
$\delta$ is defined as
\begin{eqnarray}
\label{relative value} \delta =
\frac{\sigma_{LED}-\sigma_{SM}}{\sigma_{SM}}.
\end{eqnarray}
In Figs.7(a), (b) and (c) we take $m_h$ = 115, 150 and 200 GeV,
separately. We can see that the relative value $\delta$ in J=2,
$P_{\gamma}=0.9$ polarized collision mode is much larger than in a
unpolarized collision mode, especially in the case with large
$\sqrt{s}$.

\par
In Fig.8, we present the differential cross section $d \sigma/d
\cos \theta$ of the process \rrtth as a function of $\cos \theta$,
where the scattering angle $\theta$ is the angle between Higgs
boson and one of the initial colliding photons. In Fig.8(a) the
Higgs boson mass $m_h$, $\gamma \gamma$ colliding energy
$\sqrt{s}$ and $\Lambda_T$ are fixed to be 115 GeV, 1 TeV and 2.5
TeV, respectively, and in Fig.8(b) the Higgs boson mass $m_h$,
$\gamma \gamma$ colliding energy $\sqrt{s}$ and $\Lambda_T$ are
fixed to be 115 GeV, 3 TeV and 5.5 TeV, respectively. As shown in
these figures, the differential cross sections in the SM and LED
model are the same in the J = 0 polarized $\gamma \gamma$
collision mode, and the line shapes for J = 0 and J = 2 polarized
photon collisions are quite different. Fig.8(a) shows that for the
J = 0 polarized $\gamma \gamma$ collision mode, the differential
cross section is quite large when $\theta$ is in the vicinities of
0 or $\pi$. That means the outgoing Higgs bosons in the  J = 0
polarized photon collision mode are almost collinear to the
incoming photon beams. But in the case of J = 2,
$\Lambda_T=5.5~TeV$ and $\sqrt{s}=3~TeV$ there is a large fraction
of Higgs bosons emitting at large scattering angle ($\theta \sim
\pi/2$) as shown in Fig.8(b).

\par
In Fig.9, we show the differential cross section $d \sigma/d p_t$
of the process \rrtth as a function of the transverse momentum of
the final Higgs boson $p_t$ in the range of 0 to 1 TeV, with $m_h
= 115~GeV$ and $\sqrt{s} = 2~ TeV$. As shown in this figure, the
SM and LED model results for J=0 collision mode are the same
(depicted as a dash-dotted curve), since the contribution of the
virtual graviton exchange diagrams vanishes. We can see also that
the effects of the large extra dimensions on differential cross
section $d\sigma/dp_t$ in a J = 2 polarized $\gamma \gamma$
collision mode is quantitatively enhanced comparing with that in
unpolarized collision mode. Here we can conclude that the
differential cross sections of \rrtth process in unpolarized and
$J=2$ polarized photon collision modes are sensitive to graviton
exchanges, but there is no influence of the graviton exchange in
the differential cross section of the \rrtth process with $J=0$
polarizations of initial photons.

\subsection{\eetth process}
\par
For the process \eetth, we consider the collisions in the
unpolarized and $+$$+$, $+$$-$, $-$$+$ and $-$$-$ polarized
$e^+e^-$ collision modes, the notation $+$$-$ represents the
helicities of the initial positron and electron being
$\lambda_1=+1/2$ and $\lambda_2=-1/2$, separately. In Fig.10, we
present the cross section of \eetth as a function of the c.m.s.
energy $\sqrt{s}$ with different kinds of polarizations of initial
positron/electron. Here we take $m_h=115~GeV$ and
$\Lambda_T=3.5~TeV$. Since the cross sections of the $+$$+$ and
$-$$-$ polarized modes are too small, we only present the cross
sections of unpolarized, $+$$-$ and $-$$+$ polarized $e^+e^-$
collision modes. As showed in this figure, in the region of
$\sqrt{s} < 1.5 TeV$, the curves for LED model and SM merged
together, but when $\sqrt{s} > 1.5 TeV$, each of the three curves
(solid, dashed and dotted curves) branches off into two curves.
The lower one is for the $\sigma_{SM}$ and the upper one is for
the $\sigma_{LED}$. The figure shows that the $-$$+$ polarized
$e^+e^-$ collision to produce $t\bar th^0$ would be the best
channel in probing LED effects among all the \eetth channels with
other polarizations of initial particles, if the incoming positron
and electron are completely polarized.

\par
In Table.2, we present the relative discrepancy $\delta$ defined
as in Eq.(\ref{relative value}) for different polarization
collision modes with some typical values of $\sqrt{s}$. Here we
take $m_h=115~ GeV$ and $\Lambda_T=1.5~TeV$, the polarization for
positron is taken as $P_{e^+}=0.6$ and for electron $P_{e^-}=0.8$.
We can see again the $-$$+$ polarized $e^+e^-$ collision mode is
the most significant one in probing the LED effects among all the
polarization modes in the process \eetth. So in the following
discussion, we compare the numerical results of process \eetth in
unpolarized collision mode only with that in the $-$$+$ polarized
$e^+e^-$ collision mode with $P_{e^+}=0.6$ and $P_{e^-}=0.8$.

\begin{table}[htbp]
\begin{center}
\begin{tabular}{c|c|c|c}
\hline
$\sqrt{s}$ & \multicolumn{3}{c}{~~~~$\delta$}\\[0mm]
\cline{2-4}
\vspace*{-0ex}[TeV]&~~~~$unpol.$~~~&~~~~$+- pol.$~~~~&~~~~$-+ pol.$~~~~\\[0mm]
\hline
1& 0.00236 & 0.00245 & 0.00470 \\[0mm]
2& 0.72302 & 0.529594 & 1.15039 \\[0mm]
3& 21.6655 & 15.9341 & 33.8929 \\[0mm]
\hline
\end{tabular}
\caption{\em The relative discrepancy $\delta$ in different
collision modes for \eetth process with some typical values of
$\sqrt{s}$.}
\end{center}
\end{table}

\par
Fig.11 exhibits the cross section of the process \eetth in the SM
and LED model as a function of $\sqrt{s}$, with different values
of $\Lambda_{T}$ and $m_h$. Figs.11(a), (c) and (e) are
corresponding to the unpolarized photon collision mode and their
results are coincident with those in Ref.\cite{ED 3f2}. From
Figs.11(a), (c) and (e) we can get the the same conclusion as in
Ref.\cite{ED 3f2} that in the unpolarized $e^+e^-$ collision the
virtual exchange KK gravitons can modify the cross section of
process \eetth significantly from its SM value. Figs.11(b), (d)
and (f) are corresponding to the $-$$+$ polarized $e^+e^-$
collision mode with $P_{e^+}=0.6$ and $P_{e^-}=0.8$. Analogous to
the process \rrtth, when $\sqrt{s}$ is large enough, the LED cross
sections increase while the SM ones decrease with the increment of
$\sqrt{s}$. We find that in case of taking the same values of
$\sqrt{s}$, $m_h$ and $\Lambda_T$, within the framework of the LED
model in $-$$+$ polarized $e^+e^-$ collision mode with
$P_{e^+}=0.6$ and $P_{e^-}=0.8$, the cross section is larger than
that in unpolarized $e^+e^-$ collision, but the SM cross section
remains the same in both two kinds of polarizations.

\par
In Fig.12, we present the cross section of the process \eetth as a
function of $\Lambda_{T}$ with different values of $\sqrt{s}$ and
$m_h$. Figs.12(a),(c) and (e) are for unpolarized $e^+e^-$
collision with $m_h=115~GeV$, $m_h=150~GeV$ and $m_h=200~GeV$,
respectively. Figs.12(b), (d) and (f) are for $-$$+$,
$P_{e^-}=0.8$ and $P_{e^-}=0.6$ polarized $e^+e^-$ collision, with
$m_h=115~GeV$, $m_h=150~GeV$ and $m_h=200~GeV$, separately. The
solid, dashed and dotted curves are corresponding to $\sqrt{s}$ =
1, 2, 3 TeV, respectively. The upper curves are the LED results
while the lower straight lines are the corresponding SM results.
As we expect, all the LED results decrease to their corresponding
SM results as the increment of $\Lambda_T$.

\par
In Fig.13, we show the regions in the $\sqrt{s}-\Lambda_{T}$
parameter space, where the LED effect can and cannot be observed
from process \eetth in unpolarized photon collision mode according
to the criteria shown in Eq.(\ref{upper}) and Eq.(\ref{lower}).
Figs.13(a), (b) and (c) correspond to $m_h$ = 115, 150 and 200
GeV, respectively. Here we assume the LC integrated luminosity
${\cal L} = {\cal L}_{e^+e^-}=500 fb^{-1}$. The underside region
is excluded by the unitarity bounds.

\begin{table}[htbp]
\begin{center}
\begin{tabular}{c|c|c}
\hline
$m_h$ = 115[GeV]& \multicolumn{2}{c}{~~~~~~~~~~~~$\Lambda_{T}$~~~~~~  [GeV]}\\[0mm]
\cline{1-3}
$\sqrt{s}$&~~~~$unpol.$~~~&~~~~$-+~~ pol.$~~~~\\[0mm]
\cline{2-3}\vspace*{-0ex}[TeV]& ~~$3\sigma$~~,~~~$5\sigma$ ~~& $3\sigma$~~,~~~$5\sigma$~~~ \\
\cline{1-3} \hline
0.6& ~~1114,~~1042~~&1183,~~1105~~ \\[0mm]
1.0& ~~2135,~~1988~~&2264,~~2105~~ \\[0mm]
1.5& ~~3233,~~3003~~&3416,~~3173~~ \\[0mm]
2.0& ~~4275,~~3963~~&4515,~~4188~~ \\[0mm]
2.5& ~~5287,~~4903~~&5577,~~5169~~ \\[0mm]
3.0& ~~6268,~~5815~~&6614,~~6125~~ \\[0mm]
3.5& ~~7235,~~6703~~&7629,~~7061~~ \\[0mm]
\hline
\end{tabular}
\\
\begin{tabular}{c|c|c}
\hline
$m_h$ = 150[GeV]& \multicolumn{2}{c}{~~~~~~~~~~~~$\Lambda_{T}$~~~~~~  [GeV]}\\[0mm]
\cline{1-3}
$\sqrt{s}$&~~~~$unpol.$~~~&~~~~$-+~~ pol.$~~~~\\[0mm]
\cline{2-3}\vspace*{-0ex}[TeV]& ~~$3\sigma$~~,~~~$5\sigma$ ~~& $3\sigma$~~,~~~$5\sigma$~~~ \\
\cline{1-3} \hline
0.6& ~~999,~~933~~&1059,~~988~~ \\[0mm]
1.0& ~~2058,~~1919~~&2179,~~2029~~ \\[0mm]
1.5& ~~3176,~~2953~~&3356,~~3117~~ \\[0mm]
2.0& ~~4229,~~3927~~&4463,~~4139~~ \\[0mm]
2.5& ~~5246,~~4867~~&5531,~~5128~~ \\[0mm]
3.0& ~~6241,~~5782~~&6573,~~6089~~ \\[0mm]
3.5& ~~7205,~~6672~~&7589,~~7026~~ \\[0mm]
\hline
\end{tabular}
\\
\begin{tabular}{c|c|c}
\hline
$m_h$ = 200[GeV]& \multicolumn{2}{c}{~~~~~~~~~~~~$\Lambda_{T}$~~~~~~  [GeV]}\\[0mm]
\cline{1-3}
$\sqrt{s}$&~~~~$unpol.$~~~&~~~~$-+~~ pol.$~~~~\\[0mm]
\cline{2-3}\vspace*{-0ex}[TeV]& ~~$3\sigma$~~,~~~$5\sigma$ ~~& $3\sigma$~~,~~~$5\sigma$~~~ \\
\cline{1-3} \hline
0.6& ~~787,~~728~~&833,~~770~~ \\[0mm]
1.0& ~~1964,~~1830~~&2079,~~1936~~ \\[0mm]
1.5& ~~3111,~~2893~~&3285,~~3054~~ \\[0mm]
2.0& ~~4186,~~3888~~&4413,~~4098~~ \\[0mm]
2.5& ~~5215,~~4838~~&5498,~~5095~~ \\[0mm]
3.0& ~~6212,~~5757~~&6542,~~6061~~ \\[0mm]
3.5& ~~7192,~~6657~~&7566,~~7003~~ \\[0mm]
\hline
\end{tabular}
\caption{\em The dependence of $3\sigma$ exclusion limits and
corresponding $5\sigma$ observation limits on $\Lambda_{T}$ and
$\sqrt{s}$ for the \eetth process. The three tables are
corresponding to $m_h$=115, 150 and 200 GeV, respectively. For the
$-$$+$ polarized $e^+e^-$ collision mode, we set $P_{e^+} = 0.6$
and $P_{e^-} = 0.8$.}
\end{center}
\end{table}

\par
In Table 3, we present the $3\sigma$ exclusion limits and
corresponding $5\sigma$ observation limits for the \eetth process
with $m_h = 115,~150,~200 ~GeV$ and some typical values of
$\sqrt{s}$. Both unpolarized and $-$$+$, $P_{e^+} = 0.6$, $P_{e^-}
= 0.8$ polarized $e^+e^-$ collision modes are considered. From
this table we can find that by using the process \eetth in the
$-$$+$ polarized $e^+e^-$ collision mode, the sensitivity to the
observation LED effects is improved a little comparing with that
by using the unpolarized collision $e^+e^-$ mode.

\section{summary}
\par
In this paper, we studied the LED effects in both the \rrtth and
\eetth processes at future linear colliders. We conclude that the
virtual Kaluza-Klein (KK) graviton exchange can modify
significantly both the $\sigma$(\rrtth) and $\sigma$(\eetth) for
some polarizations of initial photons from their corresponding SM
values. Our numerical results show that when we take $m_h = 115$
GeV and $\sqrt{s} = 3.5$ TeV, the effective scale $\Lambda_T$ for
the process \rrtth can be probed up to 7.8 and 8.6 TeV in the
unpolarized and $P_{\gamma} = 0.9$, J=2 polarized $\gamma \gamma$
collision modes separately, while for the \eetth process the upper
limits of $\Lambda_T$ to observe the LED effects are $6.7~TeV$ and
7.0 TeV in the unpolarized and $P_{e^+} = 0.6$, $P_{e^-} = 0.8$,
$-$$+$ polarized $e^+e^-$ collision modes, respectively.
Therefore, the sensitivity in probing LED effects can be
significantly improved by using the \rrtth process with J=2
polarizations of colliding photons. We find that the differential
cross sections of \rrtth process in unpolarized and $J=2$
polarized photon collision modes are sensitive to graviton
exchanges too, but there is no influence of the graviton exchange
on the differential cross section of the \rrtth process with $J=0$
polarizations of initial photons. We also find that in the case of
\eetth process with $-$$+$ polarization of $e^+e^-$ colliding
beams, the upper limit of $\Lambda_T$ to be probed is not
obviously improved. Comparing all the $t\bar th^0$ associated
production channels at linear colliders with different
polarizations of initial photons(or positron/electron), we find
the process \rrtth with $J=2$ polarizations of initial photons
provides a possibility to improve significantly the sensitivity in
probing the LED effects.

\par
\vskip 5mm \noindent{\large\bf Acknowledgments:} This work was
supported in part by the National Natural Science Foundation of
China and a special fund sponsored by China Academy of Science.

\section{Appendix}
\par
In this Appendix, we present the explicit expressions of the
Feynman rules which are relevant to our calculation of the
processes \rrtth and \eetth.

\par
\begin{tabular}{r}
\epsfig{file=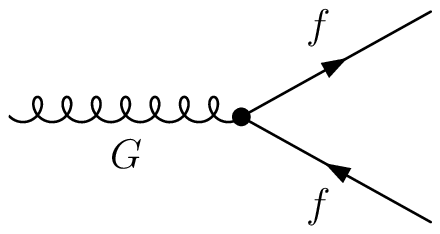, height=3 cm, width=4 cm}
\end{tabular} $C_{f f G}: -i \frac{\kappa}{8 \sqrt{V_n}}
\left[ \gamma_{\mu} (p_1 + p_2)_{\nu} + \gamma_{\nu} (p_1 +
p_2)_{\mu} - 2 g_{\mu \nu} (\rlap/{p}_1 + \rlap/{p}_2 - 2 m_{f})
\right]$  \\

\par
\begin{tabular}{r}
\epsfig{file=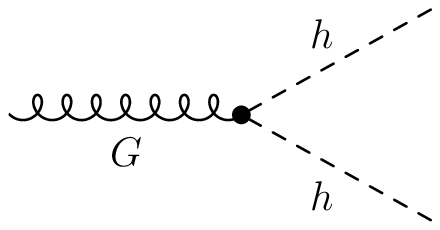, height=3 cm, width=4 cm}
\end{tabular} $C_{h h G}: i \frac{\kappa}{\sqrt{V_n}}
\left[ B_{\mu \nu \alpha \beta} p_1^{\alpha}p_2^{\beta} - A_{\mu
\nu} m_{h}^2 \right]$  \\

\par
\begin{tabular}{r}
\epsfig{file=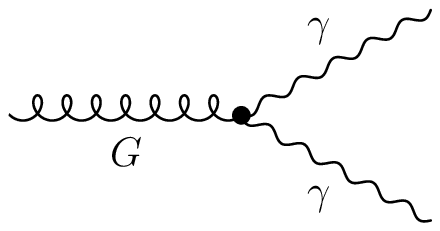, height=3 cm, width=4 cm}
\end{tabular} $C_{\gamma \gamma G}: i \frac{\kappa}{\sqrt{V_n}}
(C_{\mu \nu \alpha \beta \rho \sigma} - C_{\mu \nu \alpha \sigma
\beta \rho}) k_1^{\rho} k_2^{\sigma}
\delta^{a b}$  \\

where
\begin{eqnarray}
\label{coe}
A^{\mu \nu} & = & \frac{1}{2}g^{\mu \nu}, \nb \\
B^{\mu \nu \alpha \beta} & = & \frac{1}{2}
      (g^{\mu \nu}g^{\alpha \beta}
      -g^{\mu \alpha}g^{\nu \beta}
      -g^{\mu \beta}g^{\nu \alpha}),
       \nb\\
C^{\rho \sigma \mu \nu \alpha \beta} & = & \frac{1}{2}
       [g^{\rho \sigma}g^{\mu \nu}g^{\alpha \beta}
       -(g^{\rho \mu}g^{\sigma \nu}g^{\alpha \beta}+
         g^{\rho \nu}g^{\sigma \mu}g^{\alpha \beta}+
         g^{\rho \alpha}g^{\sigma \beta}g^{\mu \nu}+
         g^{\rho \beta}g^{\sigma \alpha}g^{\mu \nu})], \nb \\
\frac{\kappa}{\sqrt{V_n}} & = & \frac{4 \sqrt{\pi}}{M_{pl}},
\end{eqnarray}
and $(g_{\mu \nu}) = {\rm diag}\{1,-1,-1,-1\}$ is the metric
tensor of the 4-dimensional flat spacetime manifold.

\begin{flushleft} {\bf Figure Captions} \end{flushleft}

\vskip 6mm
\par{\bf Fig.1} The tree-level Feynman diagrams with graviton exchange for the process
                $\gamma \gamma \rightarrow t\bar{t} h^{0}$.\\
\par{\bf Fig.2} The tree-level Feynman diagrams with graviton exchange for the process
                \eetth.\\
\par{\bf Fig.3} (a) The cross section for the process \rrtth as a function of the c.m.s.energy
                for different kinds of collision modes. (b) The cross sections/ratio of
                $\sigma_{+-}$ and $\sigma_{++}$ as the functions of $\sqrt{s}$.\\
\par{\bf Fig.4} The dependence of the cross section for \rrtth on $\sqrt{s}$.
                (a),(c) and (e) are for unpolarized photon collisions with $m_h=115~GeV$, $m_h=150~GeV$ and
                $m_h=200~GeV$, respectively. (b), (d) and (f) are for $+$$-$, $P_{\gamma}=0.9$ polarized
                photon collisions with $m_h=115~GeV$, $m_h=150~GeV$ and $m_h=200~GeV$, respectively.\\
\par{\bf Fig.5} The dependence of the cross section for \rrtth on effective scale $\Lambda_{T}$. (a),(c) and (e)
                are for unpolarized photon collisions with $m_h=115~GeV$, $m_h=150~GeV$ and
                $m_h=200~GeV$, respectively. (b), (d) and (f) are for $+$$-$, $P_{\gamma}=0.9$
                polarized photon collisions with $m_h=115~GeV$, $m_h=150~GeV$ and
                $m_h=200~GeV$, respectively. \\
\par{\bf Fig.6} The LED effect observation area (gray) and the LED effect exclusion area
                (pale gray) for \rrtth process in the $\sqrt{s}-\Lambda_{T}$ parameter space
                considering only unpolarized beams with $m_h$ = 115, 150 and 200
                GeV, respectively.\\
\par{\bf Fig.7} The $\delta$ dependence of the process
                \rrtth with unpolarized and $J=2$, $P_{\gamma}=0.9$ polarized photon collisions on
                $\sqrt{s}$. (a),(b) and (c) are for $m_h=115~GeV$, $m_h=150~GeV$ and
                $m_h=200~GeV$, respectively.\\
\par{\bf Fig.8} The differential cross section $d \sigma/d \cos \theta$ of the
                process \rrtth as a function of $\cos\theta$ in different polarization
                collision modes (unpolarized, J = 0 and J = 2 polarized
                $\gamma \gamma$ collision modes) within both the SM and LED model. \\
\par{\bf Fig.9} The differential cross section $d \sigma/dp_t$ of the process
                \rrtth as a function of the transverse momentum of
                the final Higgs boson $p_t$ in different collision modes (unpolarized, J = 0 and J = 2
                polarized $\gamma \gamma$ collision modes) within both the SM and LED model. \\
\par{\bf Fig.10}The cross section of the process \eetth as a function of the c.m.s. energy $\sqrt{s}$ in unpolarized,
                $+$$-$ and $-$$+$ polarized $e^+e^-$ collision modes, when $m_h$=115GeV, $\Lambda_T$=3.5 TeV.\\
\par{\bf Fig.11}The dependence of the cross section for \eetth on $\sqrt{s}$. (a),(c) and (e)
                are for unpolarized $e^+e^-$ collision with $m_h=115~GeV$, $m_h=150~GeV$ and
                $m_h=200~GeV$, respectively. (b), (d) and (f)
                are for $-$$+$, $P_{e^+}=0.6$ and $P_{e^-}=0.8$ polarized $e^+e^-$ collision with
                $m_h=115~GeV$, $m_h=150~GeV$ and $m_h=200~GeV$, respectively.\\
\par{\bf Fig.12}The dependence of the cross section for \eetth on $\Lambda_{T}$. (a),(c) and (e)
                are for unpolarized $e^+e^-$ collision with $m_h=115~GeV$, $m_h=150~GeV$ and
                $m_h=200~GeV$, respectively. (b), (d) and (f) are for $-$$+$, $P_{e^+}=0.6$ and $P_{e^+}=0.8$
                polarized $e^+e^-$ collision with $m_h=115~GeV$, $m_h=150~GeV$ and $m_h=200~GeV$, respectively.\\
\par{\bf Fig.13}The LED effect observation area (gray) and the LED effect exclusion area
                (pale gray) for \eetth process in unpolarized $e^+e^-$ collision mode
                in the $\sqrt{s}-\Lambda_{T}$ parameter space. Fig.13(a), (b) and (c) are for
                $m_h$ = 115, 150 and 200 GeV, respectively.\\

\vspace*{1.5cm}
\begin{figure}[hbtp]
\vspace*{-1.5cm} \centerline{ \epsfxsize = 9cm \epsfysize = 4cm
\epsfbox{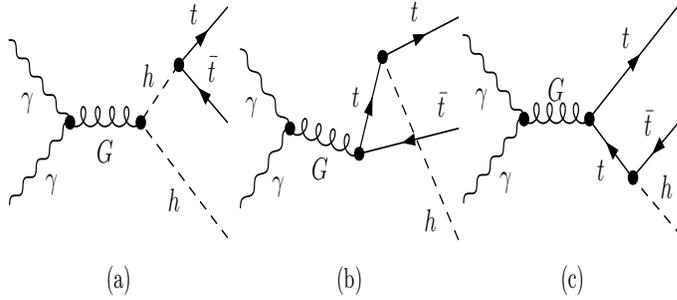}}  \vspace*{-0.5cm}\caption{The tree-level
                     Feynman diagrams with graviton exchange for the process
                     $\gamma \gamma \rightarrow t\bar{t} h^{0}$.} \label{feynrrtth}
\end{figure}

\vspace*{3.5cm}
\begin{figure}[hbtp]
\vspace*{-1.5cm} \centerline{ \epsfxsize = 9cm \epsfysize = 4cm
\epsfbox{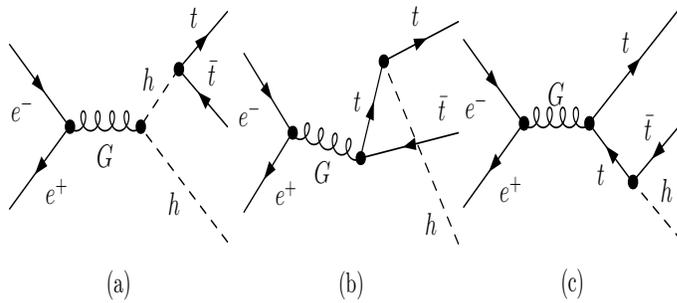}}  \vspace*{-0.5cm}\caption{The tree-level
                     Feynman diagrams with graviton exchange for the process
                     $e^+e^- \rightarrow t\bar{t} h^{0}$.} \label{feyneetth}
\end{figure}

\vspace*{3.5cm}
\begin{figure}[hbt]
\vspace*{-0.5cm} \centerline{ \epsfxsize = 15cm \epsfysize =
7cm\epsfbox{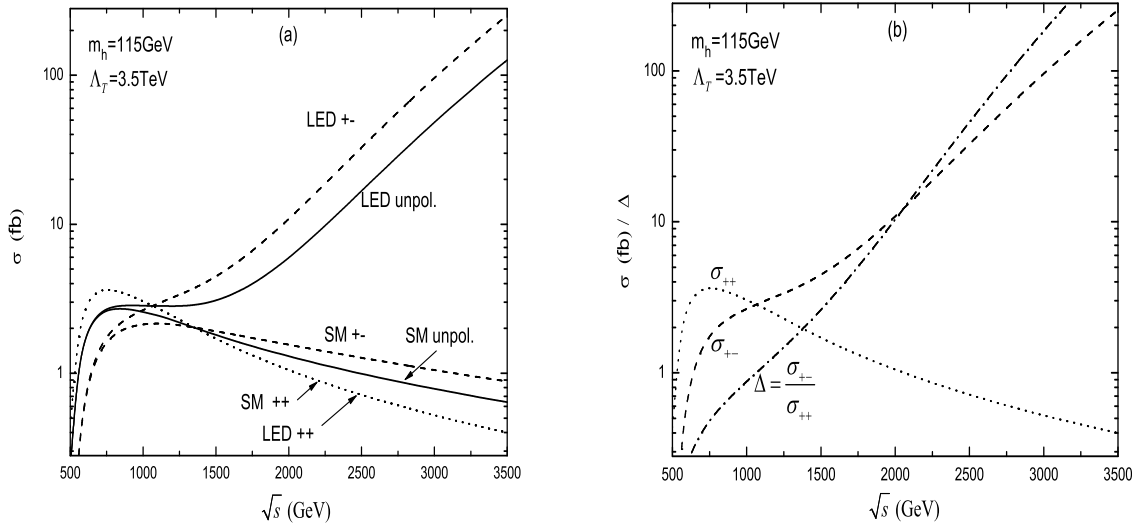}} \vspace*{-0.2cm} \caption{\em (a) The cross
                section for the process \rrtth as a function of the c.m.s.energy
                for different kinds of collision modes. (b) The cross sections/ratio of
                $\sigma_{+-}$ and $\sigma_{++}$ as the functions of $\sqrt{s}$.}
 \label{rrtthchoose}
\end{figure}

\vspace*{1.5cm}
\begin{figure}[hbtp]
\vspace*{-1.5cm}\centerline{\epsfxsize = 15cm \epsfysize =
20cm\epsfbox{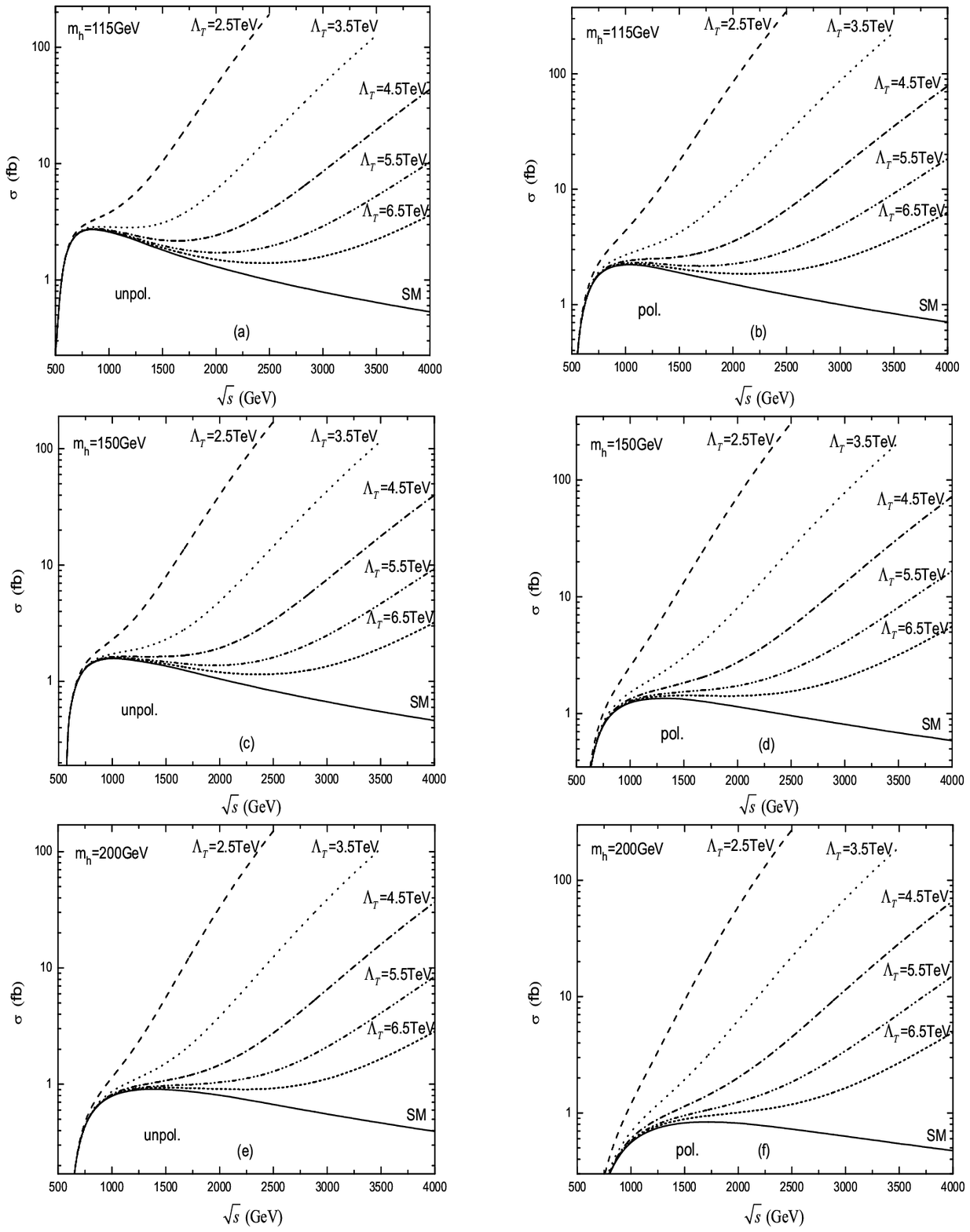}} \caption{\em The dependence of the cross
                section for \rrtth on $\sqrt{s}$.
                (a),(c) and (e) are for unpolarized photon collisions with $m_h=115~GeV$, $m_h=150~GeV$ and
                $m_h=200~GeV$, respectively. (b), (d) and (f) are for $+$$-$, $P_{\gamma}=0.9$ polarized
                photon collisions with $m_h=115~GeV$, $m_h=150~GeV$ and $m_h=200~GeV$, respectively.}
\label{rrtthcross}
\end{figure}

\vspace*{0.5cm}
\begin{figure}[hbtp]
\vspace*{-1.5cm}\centerline{\epsfxsize = 15cm \epsfysize =
20cm\epsfbox{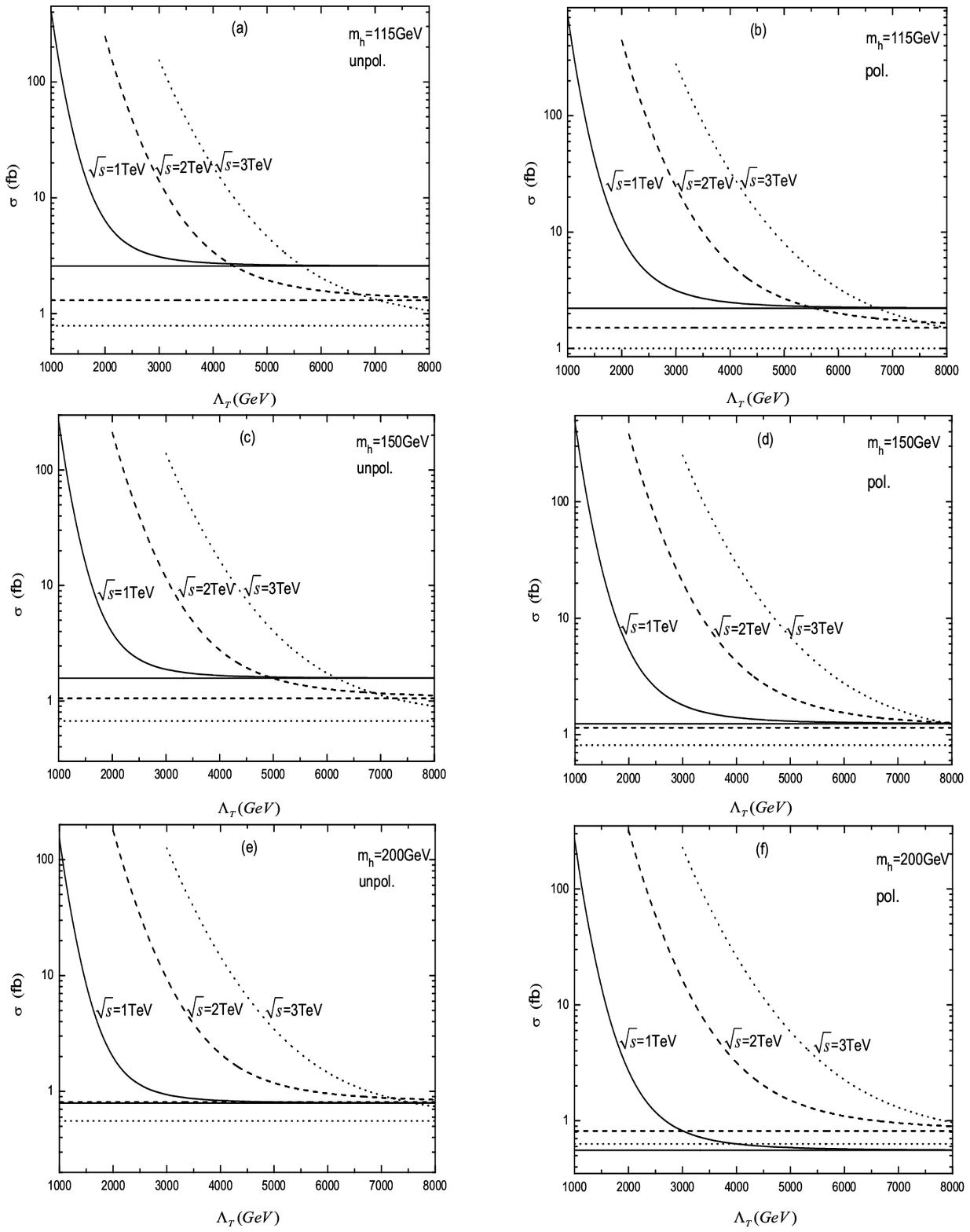}} \caption{\em The dependence of the cross
                section for \rrtth on effective scale $\Lambda_{T}$. (a),(c) and (e)
                are for unpolarized photon collisions with $m_h=115~GeV$, $m_h=150~GeV$ and
                $m_h=200~GeV$, respectively. (b), (d) and (f)
                are for $+$$-$, $P_{\gamma}=0.9$ polarized photon collisions with $m_h=115~GeV$, $m_h=150~GeV$ and
                $m_h=200~GeV$, respectively. }
\label{rrtthLamT}
\end{figure}

\vspace*{1.5cm}
\begin{figure}[hbtp]
\vspace*{-1.5cm}\centerline{\epsfxsize = 17cm \epsfysize =
5cm\epsfbox{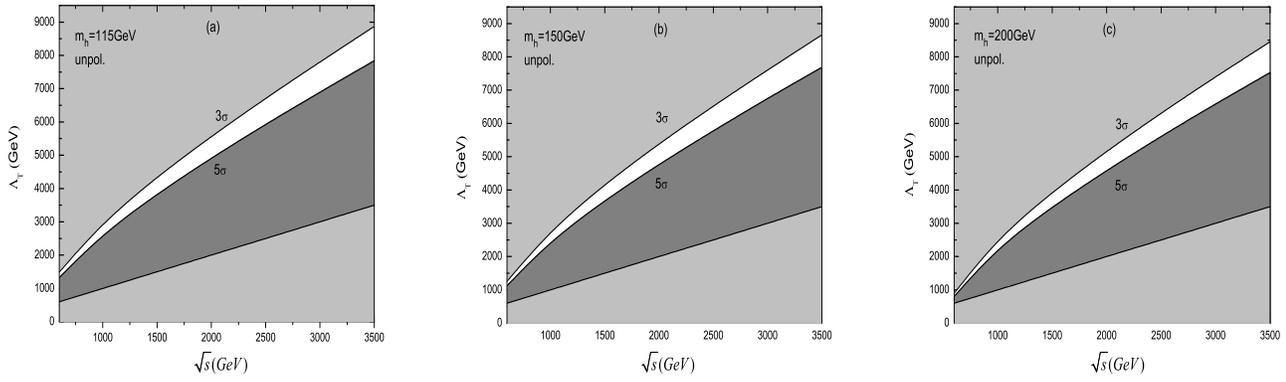}} \caption{The LED effect observation area
(gray) and the LED effect exclusion area
                (pale gray) for \rrtth process in the $\sqrt{s}-\Lambda_{T}$ parameter space
                considering only unpolarized beams with $m_h$ = 115, 150 and 200
                GeV, respectively. }
\label{rrtthsigma}
\end{figure}

\vspace*{1.5cm}
\begin{figure}[hbtp]
\vspace*{-1.5cm}\centerline{\epsfxsize = 10cm \epsfysize =
20cm\epsfbox{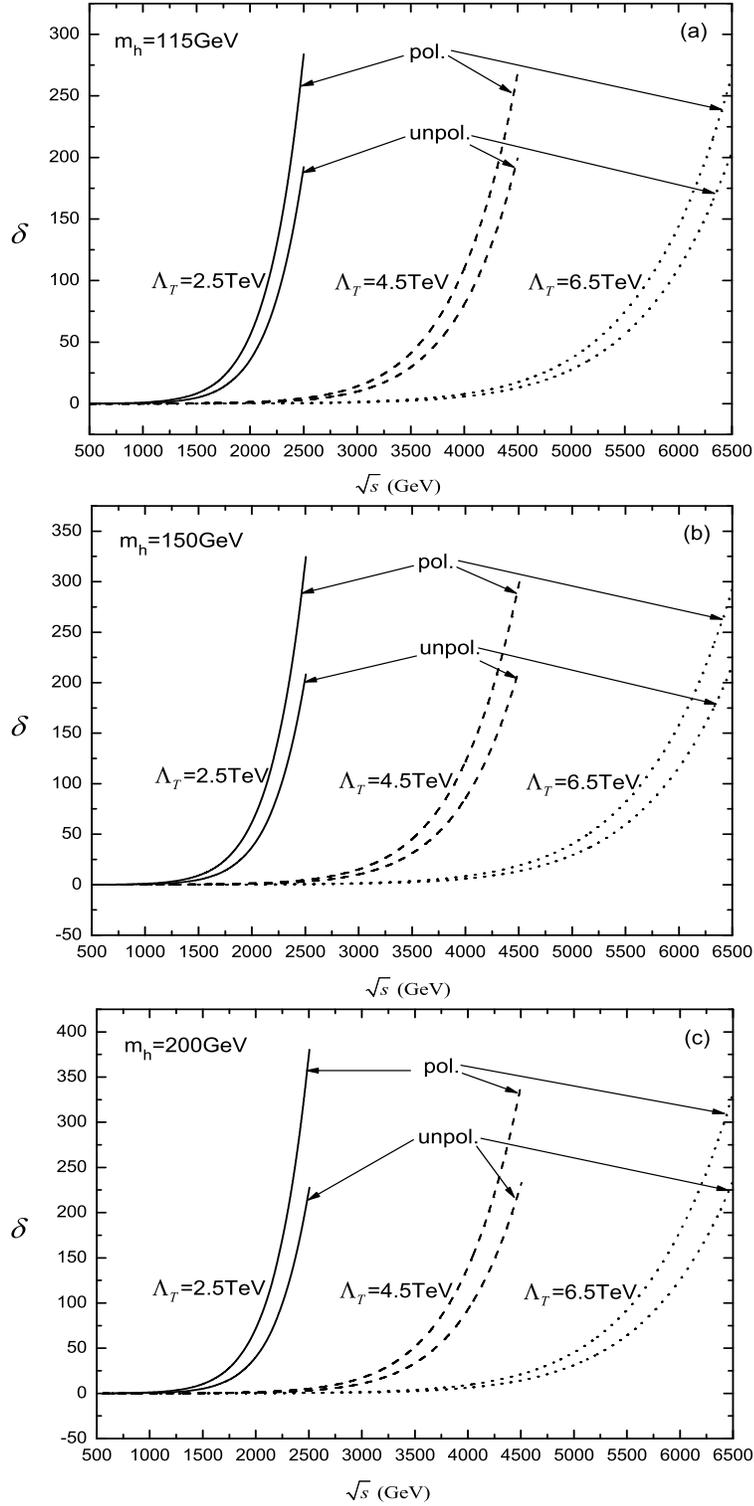}} \caption{\em The $\delta$ dependence of
                the process \rrtth with unpolarized and $J=2$, $P_{\gamma}=0.9$ polarized photon collisions on
                $\sqrt{s}$. (a),(b) and (c) are for $m_h=115~GeV$, $m_h=150~GeV$ and $m_h=200~GeV$, respectively.}
\label{rrtthdelta}
\end{figure}

\vspace*{1.5cm}
\begin{figure}[hbtp]
\vspace*{-1.5cm}\centerline{\epsfxsize = 15cm \epsfysize =
7cm\epsfbox{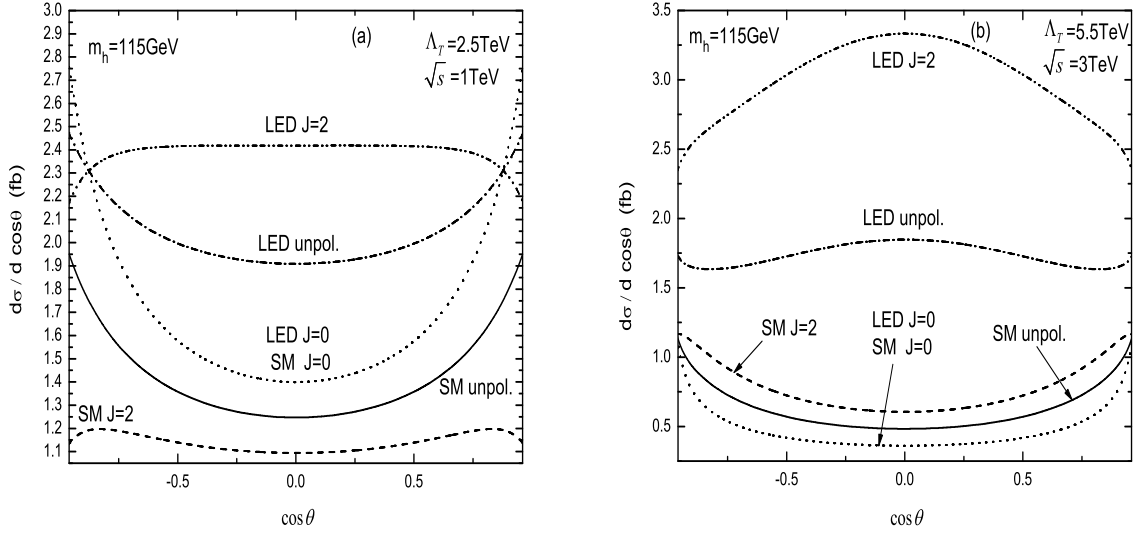}} \caption{\em The differential cross section
                $d \sigma/d \cos \theta$ of the process \rrtth as a function of $\cos\theta$ in different
                polarization collision modes (unpolarized, J = 0 and J = 2 polarized $\gamma \gamma$
                collision modes) within both the SM and LED model. }
\label{rrtthcosth}
\end{figure}

\vspace*{1.5cm}
\begin{figure}[hbtp]
\vspace*{-1.5cm}\centerline{\epsfxsize = 10cm \epsfysize =
8cm\epsfbox{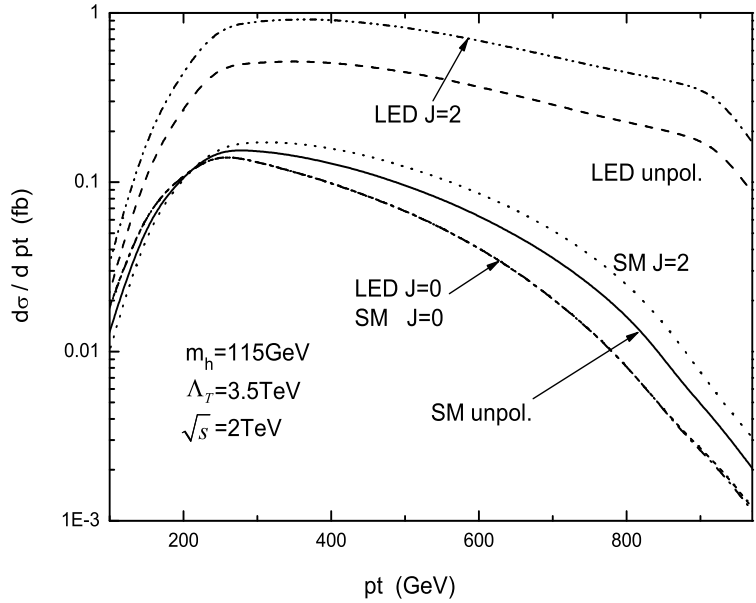}} \caption{\em The differential cross section
                $d \sigma/dp_t$ of the process \rrtth as a function of the transverse momentum of
                the final Higgs boson $p_t$ in different collision modes (unpolarized, J = 0 and J = 2
                polarized $\gamma \gamma$ collision modes) within both the SM and LED model.}
\label{rrtthpt}
\end{figure}

\vspace*{1.5cm}
\begin{figure}[hbtp]
\vspace*{-1.5cm}\centerline{\epsfxsize = 6cm \epsfysize =
6cm\epsfbox{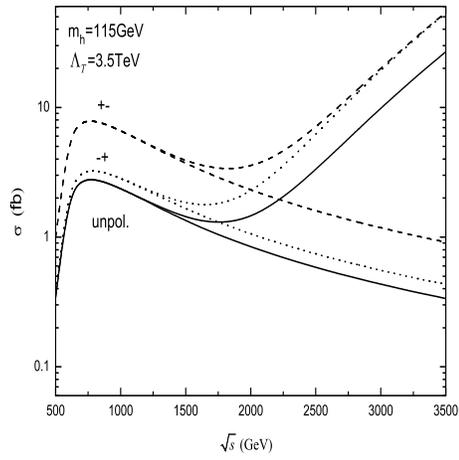}} \caption{\em The cross section of the
                process \eetth as a function of the c.m.s. energy $\sqrt{s}$ in unpolarized, $+$$-$ and $-$$+$ polarized
                $e^+e^-$ collision modes, when $m_h$=115GeV, $\Lambda_T$=3.5 TeV.}
\label{eetth_choose}
\end{figure}

\vspace*{1.5cm}
\begin{figure}[hbtp]
\vspace*{-1.5cm}\centerline{\epsfxsize = 15cm \epsfysize =
20cm\epsfbox{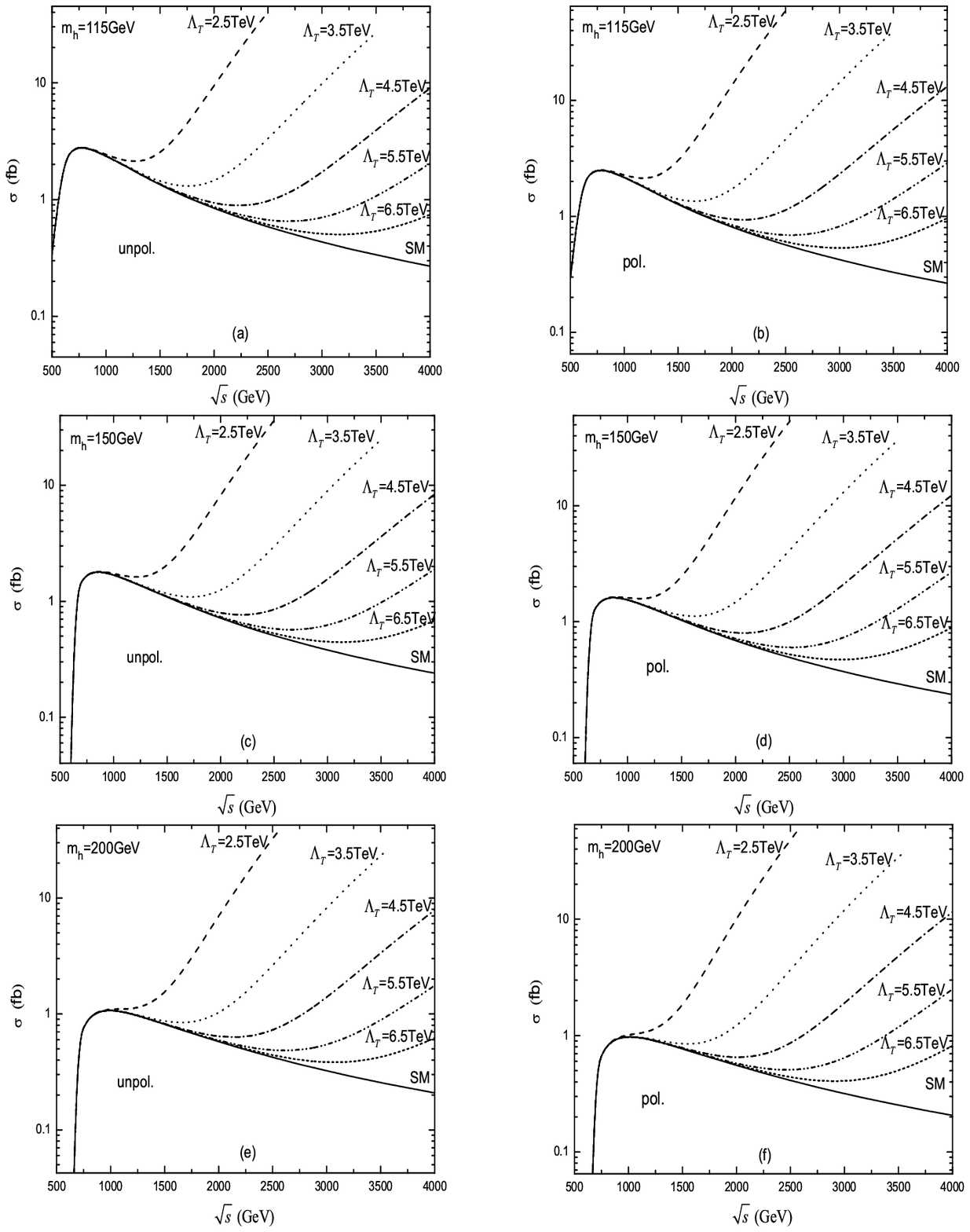}} \caption{\em The dependence of the cross
                section for \eetth on $\sqrt{s}$. (a),(c) and (e)
                are for unpolarized $e^+e^-$ collision with $m_h=115~GeV$, $m_h=150~GeV$ and
                $m_h=200~GeV$, respectively. (b), (d) and (f)
                are for $-$$+$, $P_{e^+}=0.6$ and $P_{e^-}=0.8$ polarized $e^+e^-$ collision with
                $m_h=115~GeV$, $m_h=150~GeV$ and $m_h=200~GeV$, respectively.}
\label{eetthcross}
\end{figure}

\vspace*{1.5cm}
\begin{figure}[hbtp]
\vspace*{-1.5cm}\centerline{\epsfxsize = 15cm \epsfysize =
20cm\epsfbox{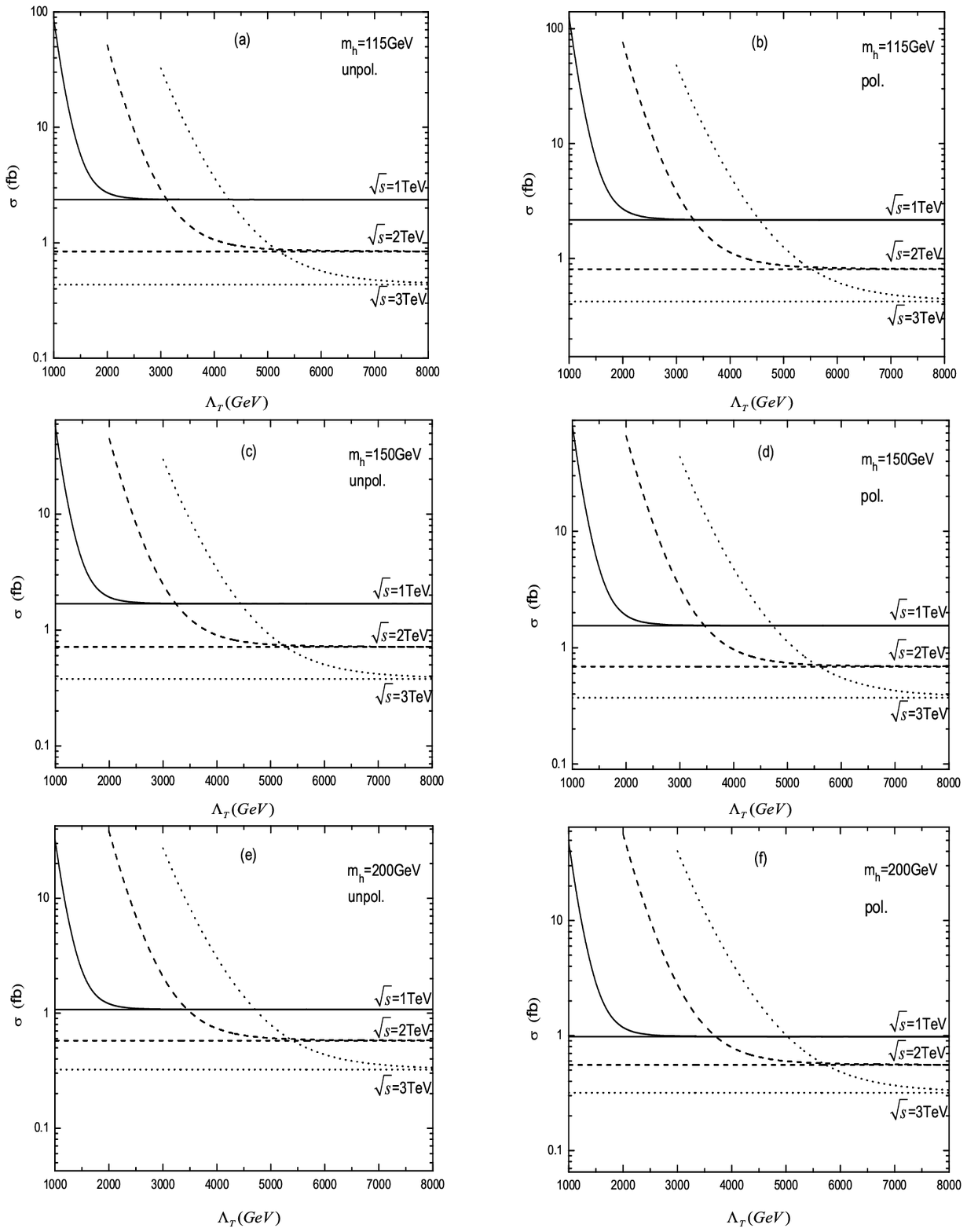}} \caption{\em The dependence of the cross
                section for \eetth on $\Lambda_{T}$. (a),(c) and (e)
                are for unpolarized $e^+e^-$ collision with $m_h=115~GeV$, $m_h=150~GeV$ and
                $m_h=200~GeV$, respectively. (b), (d) and (f)
                are for $-$$+$, $P_{e^+}=0.6$ and $P_{e^-}=0.8$ polarized $e^+e^-$ collision with
                $m_h=115~GeV$, $m_h=150~GeV$ and $m_h=200~GeV$, respectively.}
\label{eetthLamT}
\end{figure}

\vspace*{1.5cm}
\begin{figure}[hbtp]
\vspace*{-1.5cm}\centerline{\epsfxsize = 17cm
\epsfysize=5cm\epsfbox{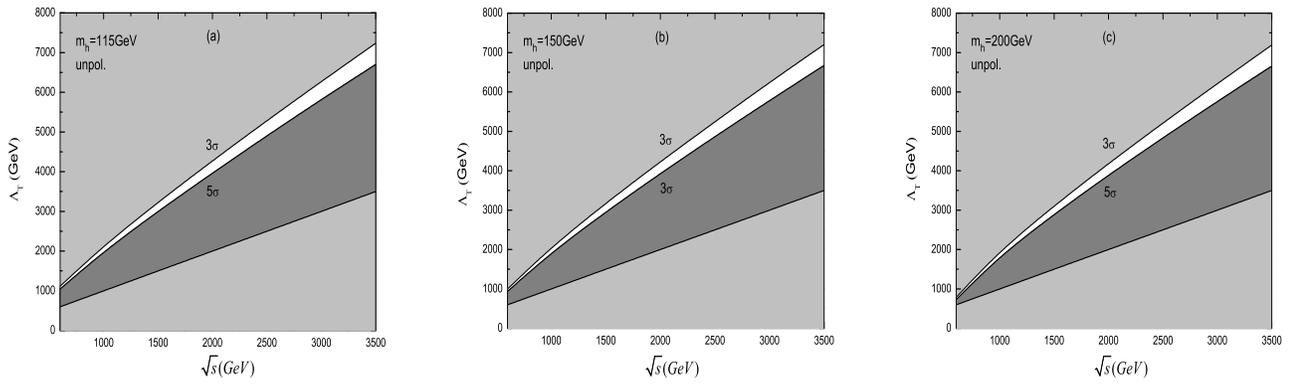}} \caption{The LED effect
                 observation area (gray) and the LED effect exclusion area
                (pale gray) for \eetth process in unpolarized $e^+e^-$ collision mode
                in the $\sqrt{s}-\Lambda_{T}$ parameter space. Fig.13(a), (b) and (c) are for
                $m_h$ = 115, 150 and 200 GeV, respectively.}
\label{eetthsigma}
\end{figure}
\end{document}